\documentclass[12pt]{iopart} 
\usepackage{graphicx,cite}

\eqnobysec

\begin{document}

\title[Bose-Einstein condensation dynamics]{Bose-Einstein condensation dynamics
in three dimensions  by the pseudospectral and finite-difference methods}

\author{Paulsamy Muruganandam$\dagger \ddagger$ 
and Sadhan K Adhikari$\dagger$}

\address{$\dagger$Instituto de F\'{\i}sica Te\'orica, Universidade
Estadual Paulista,\\
01.405-900 S\~ao Paulo, S\~ao Paulo, Brazil\\
$\ddagger$ Centre for Nonlinear
Dynamics, Department of Physics, Bharathidasan University,
\\ Tiruchirapalli
620 024, Tamil Nadu, India}

\date{\today}

\begin{abstract}

We suggest a pseudospectral method for solving the three-dimensional
time-dependent Gross-Pitaevskii (GP) equation and use it to study the
resonance dynamics of a trapped Bose-Einstein condensate induced by a
periodic variation in the atomic scattering length. When the frequency of
oscillation of the scattering length is an even multiple of one of the
trapping frequencies along the $x$, $y$, or $z$ direction, the
corresponding size of the condensate executes resonant oscillation.  
Using the concept of the differentiation matrix, the partial-differential
GP equation is reduced to a set of coupled ordinary differential equations
which is solved by a fourth-order adaptive step-size control Runge-Kutta
method. The pseudospectral method is contrasted with the finite-difference
method for the same problem, where the time evolution is performed by the
Crank-Nicholson algorithm. The latter method is illustrated to be more
suitable for a three-dimensional standing-wave optical-lattice trapping
potential.

\end{abstract}

\pacs{03.75.-b}


\noindent{Accepted in \jpb}


\section{Introduction}

The  experimental realization \cite{1} of Bose-Einstein condensates (BECs) in
dilute weakly-interacting trapped bosonic atoms  at ultra-low temperature 
initiated intense theoretical  effort to describe  the  properties  of the
condensate  \cite{11,s1,s21,s2,s31,s4,ss,s5,3y,3z,3ya,3yb}.  The
properties
of a 
condensate at zero temperature are usually described by the time-dependent,
nonlinear, mean-field Gross-Pitaevskii (GP) equation \cite{8}. The effect of
the interatomic interaction leads to a nonlinear term in the GP equation which
complicates the solution procedure. Also, to simulate the proper experimental
situation one should be prepared to deal with an anisotropic  trap \cite{k}.

A numerical study of the time-dependent GP equation is of interest, as
this can provide solution to many stationary and time-evolution
problems. The  time-independent
GP equation yields only the solution of stationary problems. As our
principal interest is in time evolution problems, we shall only  consider
the time-dependent GP equation in this paper. There are many numerical
methods
for the solution of the GP equation
\cite{s1,s21,s2,s31,s4,ss,s5,3y,3z,3ya,3yb}.

Here we suggest a pseudospectral time-iteration method
\cite{reddy,ps} for the solution of the three-dimensional GP equation with
an
anisotropic  harmonic trap and contrast it with the
finite-difference 
method \cite{dl,koo}. In the pseudospectral method the unknown wave
function is expanded in terms of a set of $N$ orthogonal polynomials. 
When this expansion is substituted
into the GP equation, the (space) differential operators operate on a set
of known polynomials and generate a differentiation matrix operating on
the unknown coefficients. Consequently, the time-dependent
partial-differential nonlinear GP equation in space and time variables is
reduced to a set of $N$ coupled ordinary differential equations (ODEs) in
time which is solved by a fourth-order adaptive step-size control
Runge-Kutta method \cite{press} using successive time iteration. In the
pseudospectral method we use the Hermite polynomials to expand the wave
function. In references \cite{3z} pseudospectral methods have been
employed for the solution of the GP equation, where a  variable step forth
order Runge-Kutta time
propagator was used, as in the present work. In  \cite{3z}
a pseudospectral  fourier-sine
basis was used for finite traps, and a corresponding complex
pseudospectral
basis  
was used for
systems
with  periodic  boundary  conditions. However, the present study seems to
be the first systematic study in dealing with the GP equation in three
space dimensions
using the pseudospectral and finite-difference approaches.  

In the finite-difference
method the time iteration
is implemented by
the split-step Crank-Nicholson scheme. The first approach  will be termed
pseudospectral-Runge-Kutta (PSRK)  method and the second
finite-difference-Crank-Nicholson (FDCN) method in the following. Here
pseudospectral and finite-difference refer to the space part and
Runge-Kutta and Crank-Nicholson refer to the time part. 
However, it should be noted that the temporal and the spatial parts in the
GP equation can  be dealt with in independent fashions. 
One can combine the pseudospectral or finite
difference approach with a specific algorithm for time stepping which can
be  Runge-Kutta  or
Crank-Nicholson  (with or without time-splitting) scheme, or any other.
In fact it
is   quite  common  to  implement some  version of the pseudospectral
method 
with  Crank-Nicholson time stepping when solving the Navier-Stokes
equation.

The PSRK method is illustrated by calculating the wave function and
energy
eigenvalue of the GP equation for different nonlinearity and trap
symmetries in three dimensions as well as by studying a resonance
dynamics. 
We find that in both cases  the
PSRK method presented here turns out to be a practical and
efficient one for the solution of the time-dependent GP equation.

Resonance is an interesting feature of an oscillation under the action of
an external periodic force manifesting in a large amplitude, when the
frequency of the external force equals a multiple of the natural frequency
of oscillation. Although, the phenomenon of resonance is
well-understood in linear problems,  in nonlinear
dynamics it is  far more nontrivial. Hence, it is worthwhile to study
the dynamics of resonance using the nonlinear GP equation. The possibility
of
generating a resonance in a BEC subject to an
oscillating trapping potential has been explored previously  \cite{6}.
Here, using the PSRK  approach we study the resonance
dynamics of a three-dimensional BEC subject to a periodic
variation in the scattering length.  One can have resonant
oscillation in the $x$, $y$, and $z$ directions, when the frequency of
oscillation of scattering length equals an even multiple of the trapping
frequency in the respective direction.  Such a
variation in the scattering length is possible near a Feshbach resonance
by manipulating an external magnetic field \cite{fesh}. There have been
recent
studies on this topic in one \cite{abd} and two
space \cite{acd}
variables.

Next we consider the FDCN  
method for the numerical solution of
the
time-dependent
GP equation 
\cite{s21,s2,s31,s4,ss}. 
In this  approach the time-dependent GP
equation is first discretized in space and time 
using a specific rule \cite{dl,koo} 
and the resultant  set of equations is  solved
by time iteration with an initial input solution \cite{s21,s2,s31,s4}. 
This procedure
leads to good result in the effective one- \cite{s21}
and two-space-variable \cite{s31,s2,s4,ss} cases
and we extend it here to the full  three-dimensional case.
In the two-space-variable case 
a three-step procedure \cite{ss} is first used to
separate  the Hamiltonian into three parts before applying 
the Crank-Nicholson scheme  in two
directions. 
In the three-space variable case we
consider a four-step procedure where the Hamiltonian is broken up into
four parts. 
We illustrate the  FDCN method in three dimensions for
the solution of the GP equation with a three-dimensional standing-wave
optical-lattice potential.

In section 2  we describe briefly the three-dimensional, time-dependent
GP
equation with an anisotropic 
trap. The PSRK and FDCN methods
 are described in section 3.  In section 4 we report the
numerical results for the wave function and energy for different 
symmetries and nonlinearities   as well as  an account of our study of
resonance 
in the anisotropic  case due to an oscillating scattering
length. In section 5 we present the solution of the GP equation for an
optical-lattice potential using the  FDCN  method and
finally, in section 6 we present a discussion of our study.

\section{Nonlinear Gross-Pitaevskii Equation}

At zero temperature, the time-dependent Bose-Einstein condensate wave
function $\Psi({\bf r};\tau)$ at position ${\bf r}$ and time $\tau $ may
be described by the following  mean-field nonlinear GP equation
\cite{11,8}
\begin{eqnarray}
\label{a} \left[ -\frac{\hbar^2\nabla ^2}{2m}
+ V({\bf r})  
+ gN_0|\Psi({\bf
r};\tau)|^2
-i\hbar\frac{\partial
}{\partial \tau} \right]\Psi({\bf r};\tau)=0. \nonumber
\end{eqnarray} 
Here $m$
is
the mass and  $N_0$ the number of atoms in the
condensate, 
 $g=4\pi \hbar^2 a/m $ the strength of interatomic interaction, with
$a$ the atomic scattering length. The normalization condition of the wave
function is
$ \int d{\bf r} |\Psi({\bf r};\tau)|^2 = 1. $

The three-dimensional trap potential is given by $  V({\bf
r}) =\frac{1}{2}m (\omega_0^2\bar  x^2+ \omega_y^2\bar 
y^2+\omega_ z^2\bar z^2)$, where
$\omega_x \equiv \omega_0$, $\omega_y$, and $\omega_z$ are the angular
frequencies in the 
$x$, $y$ and $z$ directions, respectively, and  
 ${\bf r}\equiv (\bar x,\bar y,\bar z)$ is the radial vector. The wave
function can be
written as
$\Psi({\bf r};\tau)=\psi(\bar x,\bar y,\bar z,\tau)$. 
After a transformation of variables to 
dimensionless quantities 
defined by $x =\sqrt 2 \bar x/l$, $y =\sqrt 2 \bar y/l$, $z =\sqrt 2 
\bar z/l$,  $t=\tau \omega_0, $
$l\equiv \sqrt {(\hbar/m\omega_0 )} $ and
$\phi(x, y,z,t)=\psi(\bar x,\bar y,\bar z,\tau)(
 l^3/\sqrt 8)^{1/2}$, the GP equation  becomes
\begin{eqnarray} 
\label{c}
\fl \left[
-\frac{\partial^2}{\partial x^2}
-\frac{\partial^2}{\partial y^2}
-\frac{\partial^2}{\partial z^2}
+\frac{x^2+\kappa^2y^2+\nu^2z^2}{4}
+ {\cal N}\left|
{\phi(x,y,z,t)}
\right| ^2 - i\frac{\partial }{\partial t}\right] \phi
(x,y,z,t)=0, 
\end{eqnarray} 
where ${\cal N}=8\sqrt 2 \pi n$, $\kappa=\omega_y/\omega_0$ and
$\nu=\omega_z/\omega_0$ with nonlinearity $n=N_0a/l$. The normalization
condition for the wave
function is \begin{eqnarray}\label{n1}
\int_{-\infty}^\infty dx 
\int_{-\infty}^\infty dy
\int_{-\infty}^\infty dz 
|\phi(x,y,z,t)|^2=1.
\end{eqnarray}

\section{Numerical Methods}

\subsection{Pseudospectral Runge-Kutta (PSRK) method}

First we describe the PSRK method \cite{reddy,ps} for the
one-dimensional 
GP equation in some detail and then indicate the necessary changes for the 
three-dimensional case. The  one-dimensional
GP equation is obtained by eliminating kinetic energy (derivative) terms
in $y$ and $z$, setting $\kappa = \nu =0$ and eliminating the  $y$ and
$z$ dependence of $\phi$ in (\ref{c}), e.g.,
\begin{eqnarray} \label{c2}
\left[
-\frac{\partial^2}{\partial x^2}
+\frac{x^2}{4}+{\cal N}\left|
{\phi(x,t)}
\right| ^2 -  i\frac{\partial }{\partial t}\right] \phi
(x,t)=0,
\end{eqnarray}
with  the normalization
$\int_{-\infty}^\infty dx |\phi(x,t) |^2=1.$ 

In this method the unknown function 
$\phi(x,t)\equiv \phi(x)$ is
expanded in
terms of a set of $N$ known interpolating orthogonal functions  $\{
f_j(x)\}^{N-1}_{j=0}$
as follows \cite{reddy}
\begin{equation}\label{a1}
\phi(x)\approx p_{N-1}(x)= \sum_{j=0}^{N-1}
\frac{\alpha(x)}{\alpha(x_j)} f_j(x) \phi_j, 
\end{equation}
where $\{ {x_j}\} _{j=0}^{N-1}$ is a set of distinct 
interpolation nodes,
$ \phi_j\equiv \phi(x_j)$,
$\alpha(x) $ is a weight function, and the functions 
$\{ f_j(x)\}^{N-1}_{j=0}$
satisfy $f_j(x_k)=\delta_{jk}$ (the Kronecker delta) and involve
orthogonal 
polynomials of degree $(N-1)$, so that
$
\phi(x_k)=p_{N-1}(x_k),  k=0,1,...,N-1.$
In this work  the interpolating
functions $\{ f_j(x)\}^{N-1}_{j=0}$   
are  the Hermite polynomials $H_j(x)$: $f_j(x)=H_j(x)$. 
However, one could use other polynomials, such as,
Chebyshev,  Laguerre, and Legendre. One could also consider a Fourier
(spectral) 
expansion of the wave function in terms of periodic cosine and sine
functions. 
The Hermite  polynomials are the eigenfunctions of
the  linearized  GP equation and hence 
already satisfy the boundary conditions of the 
wave function of the GP equation \cite{s1}. Consequently,
by choosing the Hermite polynomials in the expansion,
 (\ref{a1}) satisfies the proper  boundary conditions by construction.

For obtaining the numerical solution,  the GP equation (\ref{c2}) is
defined and solved on the set of grid points $x_j$. The
solution at any
other point is obtained by using the interpolation formula (\ref{a1}), or
any other convenient interpolation rule. 
The advantage of the expansion (\ref{a1}) is that when it is substituted
in the GP equation, the space derivatives operate only on the known
analytic functions $\alpha (x)$ and $f_j(x)$, so that one can define a 
matrix for the second-order space derivative \cite{reddy}:
\begin{equation}
D^2_{k,j} = \frac {d^2}{dx^2} \left[  \frac{\alpha(x)}{\alpha(x_j)}
f_j(x) \right] _{x=x_k},
\end{equation}
and the numerical differentiation process may therefore be performed as the 
matrix-vector product: $\sum_{j=0}^{N-1}D^2_{k,j}\phi_j$. 
Consequently, the partial differential equation (\ref{c2}) is reduced to a 
set of coupled ODEs in the time variable $t$
involving $\phi_j, j =0,1,...,(N-1)$. In this way we obtain a set of ODEs
by considering   the original  equations on
a  suitable set of discretization points (the roots of  Hermite
polynomials). One could  have  also used  a  Galerkin  procedure
\cite{3ya} and
projected  the equations,
essentially  by integrating  them  against  Hermite  polynomials for that
purpose. However, we do not explore this possibility in this paper.

For solving the set of ODEs we use the adaptive step-size control based on
the embedded   
Runge-Kutta formulas due to Fehlberg \cite{press}, which gives a definite
clue about
how to modify the step size in order to achieve a desired accuracy in 
a controlled way. For orders $M$ higher than four of the Runge-Kutta
formula,
evaluation of more than $M$ functions (though never more than $M+2$) is
required. This makes the classic fourth order method requiring the
evaluation of four functions  more economic. Fehlberg suggested a
fourth-order and a fifth-order method each requiring the evaluation of six
functions.  The difference between the results of these two gives the
error $\delta$ in the fourth-order method with a step size $h$, where
$\delta$ scales as $h^5:$  $\delta \propto h^5$. This scaling immediately
gives the  factor by which the step size $h$ should be reduced, so that a 
desired $\delta$  can be obtained. The detailed fourth-order and 
fifth-order Runge-Kutta formulas of Fehlberg are  given in
 \cite{press}.
We use these formulas with the  constants given by Cash and Karp
also tabulated
in  \cite{press}. For the present problem we find that the
use of Cash-Karp
constants  in the  Fehlberg formulas leads to more accurate results than
the original constants due to  Fehlberg. 

Next we specialize to the case of Hermite polynomials used to generate the
solution. 
Hermite polynomials are very convenient in this case \cite{tang}
as the
solutions
of the linear GP 
equation (\ref{c2}) with ${\cal N}=0$ are Gaussian-type functions. They
also satisfy the correct boundary conditions of the wave functions.  
In this
case $x_j$ are the roots of 
$H_{N-1}(x)$: $H_{N-1}(x_j)=0, 
j=0,1,...,N-1.$ The roots can be  found by diagonalizing a tridiagonal
symmetric Jacobi
matrix as described in
 \cite{reddy} or otherwise. 
The weight functions are taken as $\alpha(x) = \exp(-x^2/2)$,
such that expansion (\ref{a1}) becomes 
\begin{equation}
p_{N-1}(x)= \sum_{j=0}^{N-1}
\frac{\exp(-x^2/2)}{\exp(-x_j^2/2)} f_j(x) \phi_j, 
\end{equation}
where $ f_j(x)$ are taken as
\begin{equation}
f_j(x) = \frac{H_{N-1}(x)}{H_{N-1}'(x_j)(x-x_j)},
\end{equation}
where prime refers to derivative with respect to $x$. Consequently, the
differentiation matrix can be obtained from
\begin{equation}\label{d}
D^2_{k,j}=\frac{1}{\alpha (x_j)H'_{N-1}(x_j)} \frac{d^2}{dx^2}
\left[     \frac{\exp(-x^2/2)H_{N-1}(x)}{(x-x_j)}     \right]_
{x=x_k}
\end{equation}
and calculated using an algorithm described
in  \cite{reddy}.

Using the differentiation matrix (\ref{d}), the GP equation is
discretized. The grid points are 
the roots of the Hermite polynomial $H_{N-1}(x_j)=0$. However, the
actual $x_j$ values employed are obtained by scaling these roots by a
constant factor so that
most of the roots fall in the region where the
condensate wave function is sizable and only a few points are located in
the region where the wave function is negligible. 
For the spherically
symmetric case $\omega_0=\omega_y=\omega_z,$ the discretization mesh in
the three directions are identical. For the anisotropic cases, in general,
the discretization points in the three directions are different from each
other.

Though
the passage from the one-dimensional to three-dimensional PSRK
method is formally straightforward, it involves nontrivial computational
steps.  
The unknown function 
$\phi(x,y,z,t)\equiv \phi(x,y,z)$ is
expanded in
terms of a set of $N$ known interpolating orthogonal functions  $\{
f_j(x)\}^{N-1}_{j=0}$
as follows \cite{reddy}
\begin{eqnarray}\label{a3}
\fl 
\phi(x,y,z) \approx p_{N-1}(x,y,z) =
\sum_{i=0}^{N-1}
\sum_{j=0}^{N-1}
\sum_{k=0}^{N-1}
\frac{\alpha(x)}{\alpha(x_i)} 
\frac{\alpha(y)}{\alpha(y_j)} 
\frac{\alpha(z)}{\alpha(z_k)} 
f_i(x)f_j(y)f_k(z) \phi_{ijk}, 
\end{eqnarray}
where $\{ {x_j}\} _{j=0}^{N-1}$ is a set of distinct 
interpolation nodes,
$ \phi_{ijk}\equiv \phi(x_i,x_j,x_k)$,
and the  functions $\alpha$ and $f$ are defined as in the one-dimensional
case, so that $\phi(x_i,y_j,z_k)= p_{N-1}(x_i,y_j,z_k)$. The
differentiation matrices along $x$, $y$ and $z$ directions can be defined
via (\ref{d})  as in the one-dimensional case. Consequently, the
partial differential GP  equation (\ref{c})  in three space variables
is transformed to a set of ODEs in time variable on the grid points $x_i$,
$y_j$ and $z_k$, which is solved by
the adaptive step-size controlled Runge-Kutta method. The wave function at
any point is then  calculated using the interpolation formula (\ref{a3}).

\subsection{Finite-difference Crank-Nicholson (FDCN) method}

The GP equation (\ref{c}) has the form of the following nonlinear
Schr\"odinger equation
\begin{equation}\label{cx}
i\frac{\partial \phi}{\partial t} = H \phi, 
\end{equation}
where the Hamiltonian $H$ contains the different linear and nonlinear
terms including the spatial derivatives. We solve this equation by time
iteration after discretization in space and time using the finite
difference scheme \cite{dl,koo,ss}. This procedure leads to a set of
algebraic equations. 
In the present split-step method the iteration is conveniently performed
in four steps by breaking up the full Hamiltonian into different
derivative and nonderivative parts: $H=H_1+H_2+H_3+H_4$, where
\begin{eqnarray}
H_1&=&\frac{x^2+\kappa^2 y^2+ \nu ^2 z^2}{4}+ {\cal
N}|\phi(x,y,z,t)|^2,\\
H_2 &=&-\frac{\partial ^2}{\partial x^2},\quad  
H_3 =-\frac{\partial ^2}{\partial y^2}, \quad
H_4 =-\frac{\partial ^2}{\partial z^2}. 
\end{eqnarray}
The time variable is discretized as $t_n = n \Delta$ where $\Delta$ is the
time step. The solution is  advanced first over the time step $\Delta$ at
$t_n$ by solving (\ref{cx}) with $H=H_1$ to produce an intermediate
solution $\phi^{n+1/4}$ from $\phi^n$, where  $\phi^n$ is the discretized
wave function at $t_n$. This propagation is performed essentially exactly
for small $\Delta $ via 
\begin{equation}
\phi^{n+1/4} ={\cal O}_{\mbox{nd}}(H_1)\phi ^ n\equiv  \exp (-i\Delta
H_1)\phi ^ n,
\end{equation}
where ${\cal O}_{\mbox{nd}}(H_1)$ denotes time evolution with $H_1$ and
the
suffix `nd' refers to non-derivative terms. Next we perform the time
propagation corresponding to the operators $H_i, i=2,3,4$ successively via
the following semi-implicit   Crank-Nicholson schemes \cite{ss}:
\begin{eqnarray}   
\phi^{n+2/4} =  {\cal O}_{\mbox{CN}}(H_2) \phi^{n+1/4} \equiv
\frac{1-i\Delta H_2/2}{1+i\Delta H_2/2}  \phi^{n+1/4} \\ 
\phi^{n+3/4} =  {\cal O}_{\mbox{CN}}(H_3) \phi^{n+2/4} \equiv
\frac{1-i\Delta H_3/2}{1+i\Delta H_3/2}  \phi^{n+2/4} \\ 
\phi^{n+1} =  {\cal O}_{\mbox{CN}}(H_4) \phi^{n+3/4} \equiv
\frac{1-i\Delta H_4/2}{1+i\Delta H_4/2}  \phi^{n+3/4},
\end{eqnarray}
where ${\cal O}_{\mbox{CN}}(H_i)$ denotes time evolution with $H_i$ and
the
suffix `CN' refers to Crank-Nicholson. Hence the final solution at time
$t_{n+1}$ is obtained from
\begin{eqnarray} 
\phi^{n+1} =  {\cal O}_{\mbox{CN}}(H_4){\cal O}_{\mbox{CN}}(H_3){\cal
O}_{\mbox{CN}}(H_2) {\cal O}_{\mbox{nd}}(H_1) \phi^{n}. 
 \end{eqnarray}    
The details of the Crank-Nicholson discretization scheme can be found in
 \cite{ss}. 
The advantages of the above split-step method are
the following. First, the error involved in splitting the Hamiltonian is
proportional to $\Delta ^2$ and can be neglected for small $\Delta$.
A considerable fraction $H_1$ of the Hamiltonian is treated fairly
accurately without mixing with the Crank-Nicholson propagation. This
method can deal with a large nonlinear term accurately and lead to stable
and accurate converged result.

\subsection{Calculational Details}

In both the FDCN and  PSRK methods
the time iteration  is started with 
the following normalized ground-state  solution of the linear GP equation
(\ref{c}) with ${\cal N}=0$:
\begin{eqnarray}\label{ex2}
\phi(x,y,z)= \left[ \frac {\kappa \nu} {8\pi^3} \right]^{1/4}
\exp{[-(x^2+\kappa 
y ^2+\nu z ^2)/4]}.
\end{eqnarray}
 The norm of the wave function is conserved after each
iteration due to the unitarity of the time evolution operator. However, it is
of advantage to reinforce numerically  the proper normalization of the  wave
function after several (100)  time iterations 
in order to improve the precision of the result.    
Typical time step employed in the calculation is  $\Delta = 0.001$.
During the iteration the
coefficient $n=N_0a/l$ of  the nonlinear term is increased from 0 at each
step by
$\Delta_1=0.001$   until the final value
of  nonlinearity  $n$  is attained. This
corresponds to the final solution. Then several thousand  
time iterations
of the
equation were performed until a stable result is obtained.

For large nonlinearity, the Thomas-Fermi (TF) solution of the GP equation is a
better approximation to the exact result \cite{11} than the harmonic oscillator
solution (\ref{ex2}). In that case it might be  advisable to use the
TF solution as the initial trial input to the GP equation with full
nonlinearity and consider time iteration of this equation without changing the
nonlinearity. This time iteration is to be continued until a
converged solution is obtained.  However, in all the calculations reported in
this paper only   (\ref{ex2}) is used as trial input. The use of
initial TF 
solution did not lead to satisfactory  result.

\section{Numerical Results with the PSRK Method}

\subsection{Wave Function and Energy}

The present method relies on time evolution and  is suitable for both
stationary and time-evolution problems. The
stationary problems are governed by a wave function with trivial time
dependence $\phi (x,y,z,t) = \phi(x,y,z) \exp (-i\mu
t)$,  where
$\mu $ is a real energy parameter.  Thus
the stationary wave function $\phi $ and the parametric energy
$\mu$ (the chemical potential)  can be extracted from the
evolution of the
time-dependent GP equation over a macroscopic interval of
time \cite{s21,s2}.  
Here we present results for the chemical potential of 
several BECs in three dimensions for spherically symmetric, axially
symmetric and anisotropic cases. These traps with different
geometries
for $^{23}$Na have been employed in experiments as well as in a
time-independent
solution of the three-dimensional GP equation \cite{s5}. The completely
anisotropic
trap employs the parameters $\omega_0^A=354 \pi$ rad/s, $\kappa = \sqrt 2,
\nu =2$ as in the experiment of Kozuma {\it et al.} \cite{k}. The
cylindrically
symmetric trap parameters are $\omega_0^C=33.86 \pi$ rad/s, $\kappa =
\nu =13.585$  as in  \cite{ket}. The spherically symmetric trap
parameters 
are  $\omega_0^S=87$ rad/s, $\kappa =
\nu =1$ \cite{hau}. We employ the scattering length of $a=52
a_0$ of Na, where $a_0$ is the Bohr radius \cite{11}. 

\begin{table}[!ht]
\caption{The chemical potential $\mu$ corresponding to
spherical, cylindrical and anisotropical geometries, respectively, for
various numbers of condensate atoms $N_0=2^q$: 
present ($\dagger$),  
reference   \cite{s5} $(\ddagger)$.  } \begin{center}
\begin{tabular}{r|rrr|rrr|rrr}
\hline
\hline
 & \multicolumn{3}{c|}{Spherical } &\multicolumn{3}{c|}{Cylindrical}
    & \multicolumn{3}{c}{Anisotropical} \\
 \multicolumn{1}{c|}{$q$}   & \multicolumn{1}{c}{$n$} & 
\multicolumn{1}{c}{$\mu^{\mbox{\tiny
}}\dagger$}  & \multicolumn{1}{c|}{$\mu^{\mbox{\tiny
}}\ddagger$}& \multicolumn{1}{c}{$n$}&
\multicolumn{1}{c}{$\mu^{\mbox{\tiny
}}\dagger$}
& 
\multicolumn{1}{c|}{$\mu^{\mbox{\tiny
}}\ddagger$}
    & \multicolumn{1}{c}{$n$} 
& \multicolumn{1}{c}{$\mu^{\mbox{\tiny
}}\dagger$} & \multicolumn{1}{c}{$\mu^{\mbox{\tiny
}}\ddagger$}
 \\
\hline
 10 &    0.50  & 1.82 & 1.825  & 0.55  & 17.39 &17.384
    &    1.77  & 3.55  & 3.572\\
 11 &    1.00  & 2.05 & 2.065 &  1.10  & 19.32  &19.392
    &    3.54  & 4.32 & 4.345\\
 12 &    1.98  & 2.42 &  2.435  &2.19  &22.27
 & 22.359    &7.09  & 5.39  &5.425\\
 13 &    3.97  & 2.95 &   2.970 &4.38  &
  26.66  &   26.620&14.18  & 6.86  & 6.904\\
 14 &    7.93  & 3.69 &  3.719  &8.77  &32.41
    & 32.682  &28.35  & 8.83  &8.900\\
 15 &   15.86  & 4.71 &  4.743 &17.54  &41.61 &41.055    &
56.70
 & 11.55 &11.572\\
\hline
\hline
\end{tabular}
\end{center}
\end{table}

In all  calculations reported in this section we used 21 Hermite
polynomials each in $x$, $y$ and $z$ directions so that we shall be
dealing with a wave function in the form of a cubic array of
dimension $21 \times 21 \times 21$. The
maxima of  $x$, $y$ and $z$ in  discretization were chosen consistent
with the trap parameters. Typical maxima $|x|_{\mbox{max}},
|y|_{\mbox{max}},$ and $|z|_{\mbox{max}}$
are of the order of 8 for the spherical and anisotropical cases, although
a smaller $|y|_{\mbox{max}}$ and $|z|_{\mbox{max}}$ ($\sim 3$) together
with a larger $|x|_{\mbox{max}}$ $(\sim 10)$  
have been used in
the axially symmetric case because of large distortion of trap parameters 
in that case ($\kappa=\nu = 13.585$).

In table 1 we list the chemical potentials for different symmetries
obtained
from the PSRK approach  as a function of number of
atoms $N_0=2^q$ in the condensate and compare them with those of a calculation
based on a  discrete variable representation of the time-independent GP
equation \cite{s5}. In the present  calculation    the chemical potentials 
are  extracted from results of time evolution of the GP equation. The results 
were evaluated at a space point near the center of the BEC and averaged over
several samples of calculation.  The error (standard deviation) of the time
averaged chemical potential is typically of the order of $0.2 \%$.  Considering
that the present approach is based on time evolution, the precision is quite
satisfactory $-$ about less than a percentage point of discrepancy when
compared to results of   \cite{s5}. It was most difficult to obtain good
convergence in the cylindrical case with highly distorted trap ($\kappa = \nu
=13.585$). A more carefully chosen  values of  the maxima   $|x|_{\mbox{max}}$
$ (\sim 10),$  and $|y|_{\mbox{max}} =|z|_{\mbox{max}} $ $(\sim 3) $ were
needed in this case. 

\begin{table}[!ht]
\caption{The convergence of the rms sizes  $\langle
x,y,z\rangle _{\mbox{rms}}$
calculated with the PSRK method  for the
anisotropical $q=12$ case  for various number of space discretization
points $N$. 
  } \begin{center}
\begin{tabular}{rrrr}
\hline
\hline
$N$ & $\langle x\rangle _{\mbox{rms}}$  & $\langle y\rangle _{\mbox{rms}}$
& $\langle z\rangle _{\mbox{rms}}$\\
\hline
7 & $1.767\pm 0.029$ & $1.295 \pm 0.028$ & $0.981 \pm 0.015$\\
8& $1.707\pm 0.025$ & $1.277\pm 0.018$ &  $0.959\pm 0.015$\\    
9 & $1.654\pm 0.022$ & $1.251\pm 0.010$& $0.957\pm 0.004$\\
10& $1.690\pm 0.015$ & $1.257\pm 0.008$ & $0.959\pm 0.006$\\
      11    &  $1.701\pm 0.002 $& $1.263\pm 0.005$ 
& $0.960\pm 0.007$\\
        12   &  $ 1.691\pm 0.015$& $1.263\pm 0.005$& 
$0.960\pm 0.007$\\
        13    &$  1.689\pm 0.004 $& $1.262\pm0.012$ 
&$ 0.961\pm 0.008$\\
        14     &$ 1.691\pm 0.003  $&    $1.262\pm 
0.013 $ & $0.962\pm 0.008$\\
        15 &  $   1.693\pm 0.006  $   & $1.263\pm 
0.013 $&  $   0.961 \pm 0.009$\\
        16  & $   1.692 \pm 0.005 $  &  $1.263\pm0.013$
 &    $ 0.962\pm 0.009$\\
\hline
\hline
\end{tabular}
\end{center}
\end{table}

In the following we study the convergence of the PSRK method in the
anisotropical case with $q=12$ calculated with $|x|_{\mbox{max}}=9$,
$|y|_{\mbox{max}} =6.5$ and $|z|_{\mbox{max}}=5$ which seems to be optimal
and were found after some experimentation. These values were taken to be
roughly five times the root mean square (rms) sizes $\langle x,y,z\rangle
_{\mbox{rms}}$ in the respective directions. Here we consider the
convergence of these rms sizes. After the proper nonlinearity is
introduced in the GP equation, if we continue to integrate using the
Runge-Kutta ODE solver routine the wave function and the rms sizes
fluctuate a little. This type of fluctuation is common to time stepping
methods for a partial differential equation.  To quantify this fluctuation
we calculate the rms sizes over 250 successive samples generated after 20
time steps of 0.01 each in the Runge-Kutta ODE routine. This corresponds
to a total time interval of $250\times 20 \times 0.01$ or 50 units. Then
we calculate the mean rms sizes and the standard deviations which we show
in table 2 for different number of expansion points $N$ in  (\ref{a1}).  
The numerical error in the rms sizes is typically less than one percent
for $N>10$ and the convergence is quite satisfactory. Considering that we
are solving a partial differential equation in four variables this error
is small. It will be difficult to obtain similar precision in the FDCN
method in this problem even with a significantly larger number of space
discretization points ($N>100$)  in each direction. The FDCN method is no
match to the PSRK method in this problem with a smooth potential.  
However, the FDCN method seems to be very suitable for a rapidly varying
potential, such as the optical-lattice potential considered in section 5,
which requires a large number of equally distributed spatial
discretization points for a faithful reproduction of the potential.
Although, the present PSRK method yields satisfactory result for the
stationary problem, its main advantage lies in its ability to tackle
time-dependent problems as we shall see in the following.

\subsection{Resonance Dynamics}

The appearance of a resonance in the oscillation of a BEC due to a periodic
variation of the scattering length has been postulated recently in spherically
\cite{abd}  and axially symmetric traps \cite{acd}.  Here we extend this
investigation to the more realistic and complicated case of an anisotropic 
trap.

\begin{figure}[!ht]
\begin{center}
\includegraphics[width=0.7\linewidth]{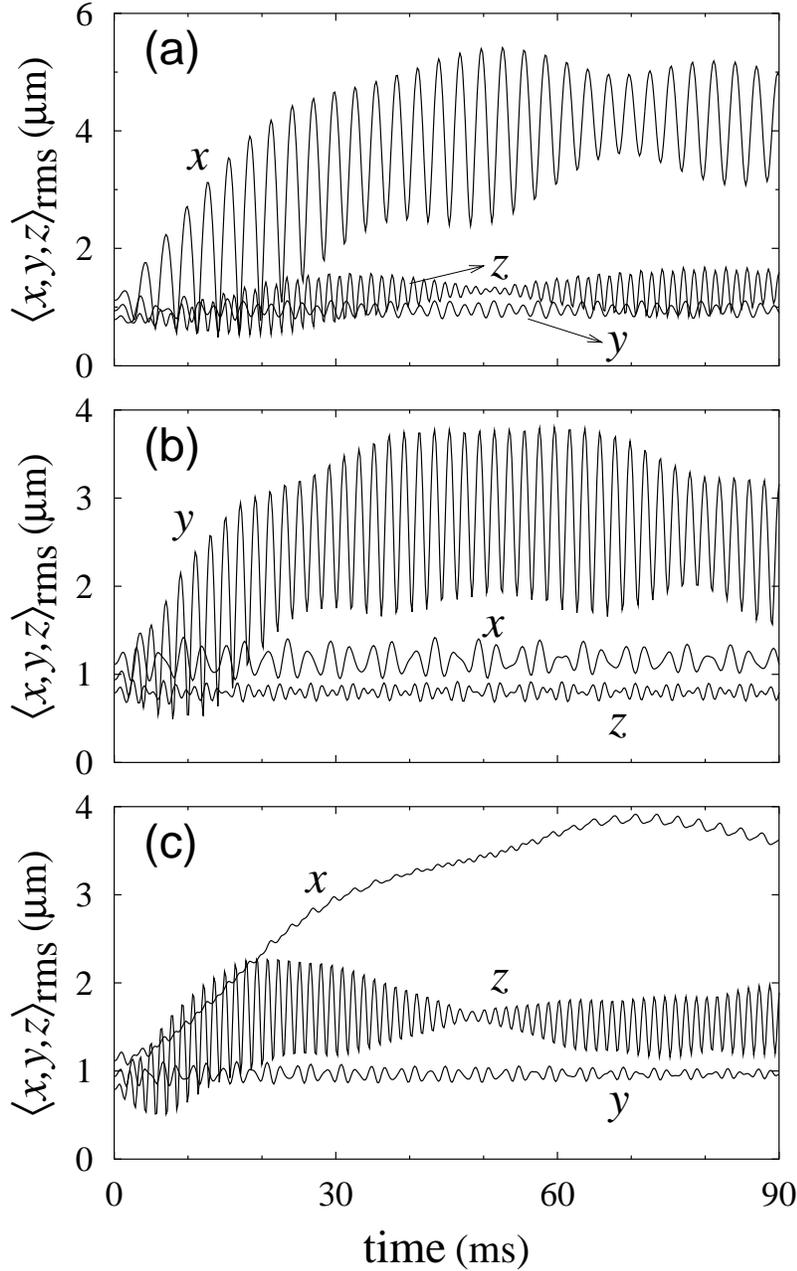}
\end{center}
\caption{The   rms sizes $\langle x\rangle _{\mbox{rms}}$,
$\langle y\rangle _{\mbox{rms}}$, and
$\langle z\rangle _{\mbox{rms}}$ vs. time  for a BEC in a
harmonic trap with $\kappa =\sqrt 2, \nu =2$ subject to a 
sinusoidal variation of 
nonlinearity $n=0.3 [\sin (\Omega t)$ with 
(a) $\Omega =2$,
(b) $\Omega =2\sqrt 2$, and (c)  $\Omega =4$.}
\end{figure}

In the case of a damped classical oscillator under the action of an
external periodic force  a
resonance appears when the
frequency of the external force is equal to or a  multiple of the natural 
frequency of oscillation of the system. 
The natural
frequency of oscillation of a trapped three-dimensional BEC along  a
certain direction ($x$, $y$, or $z$) is twice the trapping frequency 
in that direction \cite{ss,fb2}. For  a trapped BEC
subject to a nonlinear external force due to a periodic variation in the
scattering length  given by $n=n_0 \sin (\Omega t)$,  
in analogy with the  damped classical oscillator a
resonance in the oscillation of the BEC along a particular direction 
is expected when $\Omega$ equals the natural frequency of oscillation in
that direction 
or a multiple of this frequency. In the present numerical study of
the anisotropic   case we find that this is indeed the case. 

To investigate the phenomenon of resonance we solve (\ref{c}) with
${\cal N} = 8\pi \sqrt 2 n_0 \sin (\Omega t)$ with $n_0 = 0.3$.
Actually, $n_0$ has to be less than a critical value $n_{\mbox{crit}}$
($\sim 0.55$) in order to avoid collapse for attractive interaction
\cite{11,s4}. The
actual critical value depends on the asymmetry parameters $\kappa$ and
$\nu$ of the harmonic trap. For the spherically symmetric case $\kappa
=\nu =1$, $n_{\mbox{crit}}=$ 0.575 \cite{11}.  To study the resonance dynamics   we use 31
Hermite
polynomials in each of $x$, $y$, and $z$ directions. The maximum values of
$|x|$, $|y|$, and $|z|$ employed in space discretization are each 15.
The resonance is best studied by plotting the
rms  sizes $\langle x,y,z\rangle _{\mbox{rms}}$
vs. time. 
In this study we employ the trap parameters $ \omega_x\equiv \omega_0
=354 \pi$ rad/s,
$\kappa= \sqrt 2$ and $\nu =2$ \cite{s5,k}, so that the unit of
time is $\omega_0 ^{-1}=0.9$ ms and of length is $l/\sqrt 2 =1.115$
$\mu$m. For values
of $\Omega$ off the resonance 
the rms sizes exhibit oscillation of very small amplitude. For
resonance frequencies $\Omega$, the rms sizes  execute oscillation
of large amplitude. 

To  illustrate the resonance we plot in figures 1 (a), (b), and
(c) $\langle  x,y,z\rangle _{\mbox{rms}}$ vs. time for $\Omega = 2$,
$2\sqrt 2$,
and 4,
respectively. These values of $\Omega$ correspond to the natural frequency
of oscillation of the condensate along the $x$, $y$, and $z$ directions,
respectively  \cite{ss,fb2}.  In addition,  $\Omega = 4$ is also twice the
natural
frequency
of oscillation of the condensate along the $x$ direction. So for  $\Omega
= 2$ and  $2\sqrt 2$ the rms values of $x$ and $y$ execute resonance
oscillation as shown in figures 1 (a) and (b), respectively. At resonance in
a particular direction, the corresponding dimension increases with time in
a oscillatory fashion. The rms values
of the other components do not show resonance. 
For $\Omega   =4$, rms
values of both $x$ and $z$ exhibit resonance. 
In case of
$ \langle z\rangle _{\mbox{rms}}$
this corresponds to the lowest harmonic and for  $ \langle x\rangle
_{\mbox{rms}}$
this
corresponds to the first excited state. However, the behavior of  $
\langle
x\rangle
_{\mbox{rms}}$ and  $ \langle z\rangle
_{\mbox{rms}}$ are different in this case.

\section{Numerical Results with the FDCN Method}

For a smooth trapping potential the PSRK method discussed in the
last section yields excellent result with a smaller number of spatial
discretization points which are unevenly distributed (at the
scaled roots of the Hermite polynomial) compared to the
FDCN method employing a relatively large number of evenly
distributed spatial discretization points. For smooth potentials the
CPU time in the  PSRK method  could be even an order of
magnitude smaller than that in  the FDCN method.   
However, the  FDCN
method has advantage in the case of a rapidly varying trapping potential 
requiring a large number of evenly distributed spatial discretization
points for a proper description of the trapping potential. One such
potential is the optical-lattice trapping potential recently
used in BEC experiments in one \cite{cata}  and three \cite{greiner}
dimensions.

The optical-lattice  potential created with the standing-wave laser field
of
wavelength $\lambda$ is given by $V_{\mbox{opt}}=V_0E_R\sum _{i=1}^3\sin^2
(k_Lx_i)$, with $E_R=\hbar^2k_L^2/(2m)$, $k_L=2\pi/\lambda$, and $V_0$ the
dimensionless strength of the optical-lattice  potential governed by 
the intensity of the laser \cite{greiner}. 
In terms of the
dimensionless laser wave
length $\lambda _0= \sqrt2\lambda/l $ and  the dimensionless
standing-wave energy parameter $E_R/(\hbar \omega)= 4\pi^2/\lambda
_0^2$, $ V_{\mbox{opt}}$ is given by
\begin{equation}\label{pot}
\frac{ V_{\mbox{opt}}}{\hbar \omega}=V_0\frac{4\pi^2}{\lambda_0^2}
\sum_{i=1}^3 \sin ^2 \left(
\frac{2\pi}{\lambda_0}x_i
\right).
\end{equation}
In the actual experiment \cite{greiner} this dimensionless standing-wave
optical-lattice potential is superposed on the harmonic trapping potential
of (\ref{c}) and we present the solution of (\ref{c})
under the action of spherically symmetric  harmonic ($\kappa =\nu =1$) as
well as the optical-lattice  potential
(\ref{pot}). 

\begin{figure}
\begin{center}
\includegraphics[width=0.8\linewidth]{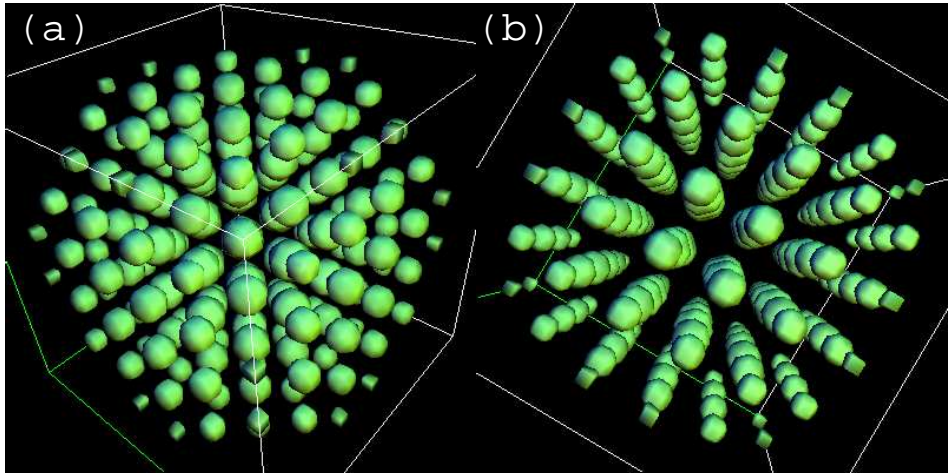}
\end{center}
\caption{Three-dimensional contour plot of the interior part of the 
BEC ground state wave
function under the
combined action of the harmonic and the optical trap for
$n\equiv N_0a/l=10$, $\lambda _0 =1$ and $V_0=10$ on a cubic lattice of
size
$3\times 3 \times 3$ ($-1.5<x,y,z<1.5$): (a) view along the diagonal of
the 
cube and
(b) along one of the axes of the cube.  
}\end{figure}

To calculate the wave function we discretize the GP equation with time step
0.001 and space step 0.1 along $x$, $y$, and $z$ directions spanning the cubic
region $-4 <x,y,z <4$. Consequently, the wave function is defined on  a cubic
array of dimension $80 \times 80 \times 80$. Starting the time iteration with
the known harmonic oscillator solution for nonlinearity ${n =0}$ and $V_0=0$,
the nonlinearity $n $  and the optical-lattice strength $V_0$ are slowly
increased by equal amount in $1000n$ steps of time iteration until the desired
values of $n$ and $V_0$ are obtained. Then without changing any parameters the
solution is iterated several thousand times so that a converged solution
independent of the initial inputs is obtained. To illustrate the present method
we calculate the wave function for the ground state  for nonlinearity $n\equiv
N_0a/l = 10$, laser wave length $\lambda _0 =1$ and optical-lattice strength
$V_0=10$. Two views of the three-dimensional contour plot of the central
part of
the  wave function on a cubic lattice of size $3 \times 3 \times 3$ are
shown
in figures 2 (a) and (b)  as seen from two different angles.  The droplets of
BEC at each pit of the optical-lattice potential can be identified in figures
2. There are about 10 occupied cites in each of $x$, $y$, and $z$
directions of which the central part is shown in figure 2.

\section{Conclusion}

In this paper we propose and implement a PSRK method
\cite{reddy} and
contrast it with the  
 FDCN method \cite{ss} for the
numerical solution of the time-dependent nonlinear GP equation under the
action of a three-dimensional trap by time iteration. 
In the PSRK method the unknown wave function
is expanded in a set of known polynomials (Hermite). 
Consequently, the partial differential GP equation in
three space and time variables becomes a set of coupled ODEs in time for
the unknown coefficients, which is solved by a fourth-order adaptive
step-size control
Runge-Kutta method \cite{press}. In the  FDCN
method 
the full Hamiltonian is split into the derivative
and nonderivative parts. In this fashion the time propagation with the
nonderivative parts can be treated very accurately. The time derivative
part is  treated by the Crank-Nicholson scheme  in three 
independent steps. Both  methods lead to 
stable and accurate results. The final result remains stable for thousands
of time iteration of the GP equation.

We applied the PSRK  method for the numerical study of certain
stationary and
time-evolution problems.  We solved the GP equation for spherically
symmetric,
axially symmetric and anisotropic  cases and calculated the chemical
potential 
for
different nonlinearities. The results  compare well with those
obtained by a time-independent approach \cite{s5}. The two sets of 
energy values agree to within a fraction of a percentage point. 
The numerical error in the method is found to be less than one percent
with a small number of expansion functions ($N\sim 20$).
The
PSRK method was also used to study the resonance dynamics of an
anisotropic BEC under the action of periodic sinusoidal variation of
the scattering length. When this period of oscillation coincides with the
natural frequency of oscillation of the BEC along $x$, $y$, or $z$
directions or a multiple thereof the corresponding rms size executes
resonant oscillation \cite{abd,acd}.
Using the FDCN method we study the ground state of a
BEC under the combined action of a harmonic and a periodic
optical-lattice trapping potential in three space dimensions.

The domain of  applicability of the 
PSRK   and FDCN methods seems to be
complementary rather than overlapping.  The PSRK  method is more
efficient and economic for smooth potentials where a relatively small
number of expansion functions and unevenly distributed
spatial discretization 
points seems to be adequate. The FDCN  method employs 
a large number of evenly distributed  
spatial discretization
points and is suitable for a rapidly varying potential, such as the
optical-lattice potential, favoring such a distribution. 

\ack
The work is supported in part by the Conselho Nacional de Desenvolvimento
Cient\'\i fico e Tecnol\'ogico and Funda\c c\~ao de Amparo \`a Pesquisa do
Estado de S\~ao Paulo of Brazil. Part of the work of PM is  supported by
the Department of Science and Technology, Government
of India.

\end{document}